# Piezoelectric Strain FET (PeFET) based Non-Volatile Memories

Niharika Thakuria, Reena Elangovan, Anand Raghunathan and Sumeet K. Gupta

*Abstract*— We propose non-volatile memory (NVM) designs based on Piezoelectric Strain FET (PeFET) utilizing a piezoelectric/ferroelectric (PE/FE such as PZT) coupled with 2D Transition Metal Dichalcogenide (2D-TMD such as $MoS_2$) transistor. The proposed NVMs store bit information in the form of polarization (*P*) of the FE/PE, use electric-field driven *P*-switching for write and employ piezoelectricity induced dynamic bandgap modulation of 2D-TMD channel for bit sensing. We analyze PeFET with COMSOL based 3D modeling showing that the circuit-driven optimization of PeFET geometry is essential to achieve effective hammer-and-nail effect and adequate bandgap modulation for NVM read. Our results show that distinguishability of binary states to up to 11X is achieved in PeFETs. We propose various flavors of PeFET NVMs, namely – (a) high density (HD) NVM featuring a compact access-transistor-less bit-cell, (b) 1T-1PeFET NVM with segmented architecture, targeted for optimized write energy and latency and (c) cross-coupled (CC) NVM offering a trade-off between area and latency. PeFET NVMs offer up to 7X smaller cell area, 66% lower write energy, 87% lower read energy and 44% faster read compared to 2D-FET SRAM. This comes at the cost of high write latency in PeFET NVMs, which can be minimized by virtue of optimized PE geometry.

*Index Terms*— Ferroelectric, non-volatile memory, piezoelectric, strain, transition metal dichalcogenides

## I. Introduction

In the current era of machine learning, the requirements for data processing and storage have exploded [1]. Along with technology scaling, increasing the number of processing cores has been a viable solution for this situation [2]. State-of-the-art processors comprise of CMOS SRAM based cache which suffer from limited scaling capability and high static leakage, leading to gaps in bridging processor-memory performance. A memory that is fast, energy efficient, highly compact and has zero standby leakage is recommended for the evolving needs.

Emerging non-volatile memories (NVM) such as spin-memories [3], resistive RAMs [4], Phase Change Memories [5], ferroelectric (FE) based memories [6]–[9] address concerns of standby leakage. However, most of them [3]-[5] rely on current-driven write, which is slow and energy-hungry. An exception to this is FE-based NVMs employing low power electric-field (*E*) driven write. However, existing FE-based designs suffer from limitations such as destructive read in FERAMs, retention/variability concerns in FEFETs and gate leakage in FE-Metal-FETs (FEMFET) [6], [8], [9]. Hence, a natural question is if an NVM can be designed to retain useful features of FE (e.g., *E*-driven write) while alleviating the said concerns. Recently, a steep switching device was proposed which employs electrostrictive or piezoelectric (PE) effect of materials such as PMN-PT in conjunction with 2D Transition Metal Dichalcogenides (2D-TMD) FETs [10] to achieve sub-60mV/decade sub-threshold swing. Motivated by the unique operation of this device, we propose an NVM device which utilizes PE/FE material (such as PZT-5H) coupled with 2D-TMD FET (such as $MoS_2$) to tackle issues in FE-based NVMs. The proposed piezoelectric strain FET (PeFET) based NVM features (a) *FE polarization (P)-based bit-storage* (b) *E-driven write and* (c) *coupling of piezoelectricity of FE with dynamic bandgap ($E_G$) tuning of 2D-TMD for read* [11]. PeFET inherits the advantages of FE and 2D-TMDs (such as scalability and enhanced gate control [10]). It leverages unique coupling of PE/FE with 2D-TMD for designing NVMs that can mitigate challenges associated with FE-based NVMs. Due to its unique structure and operational mechanism, the proposed PeFET NVM can potentially achieve (i) retention as high as FERAMs (the highest amongst all FE NVMs), (ii) non-destructive read and (iii) elimination of the gate leakage concerns due to absence of any floating metal layers. However, the benefits of PeFETs are accompanied by certain device limitations, which we tackle at the circuit/array level by pursuing a device-circuit co-design approach. With this in view, we propose multiple PeFET based NVMs. Our contributions in this manuscript are as following:

1. We present Piezoelectric Strain FET (PeFET), which utilizes *P*-based bit storage in PZT-5H, *E*-driven *P*-switching, and piezoelectricity induced dynamic bandgap modulation in 2D-TMD ($MoS_2$) channel for bit sensing.

2. We propose PeFET NVMs namely, High Density (HD), 1T-1PeFET and cross-coupled (CC) cell. HD is targeted for high density memories while the latter two for low power write.

3. We conduct an extensive device-circuit co-design to establish the design requirements and area-energy-latency tradeoffs of PeFET NVMs with respect to 2D-FET SRAMs.

This paper was submitted on January 10, 2022 and accepted on xx, 2022. This research was supported by Army Research Office (W911NF-19-1-048).

Thakuria, Elangovan, Raghunathan and Gupta are with the School of Electrical and Computer Engineering, Purdue University ({nthakuri, elangovr, raghunathan, guptask} @purdue.edu).

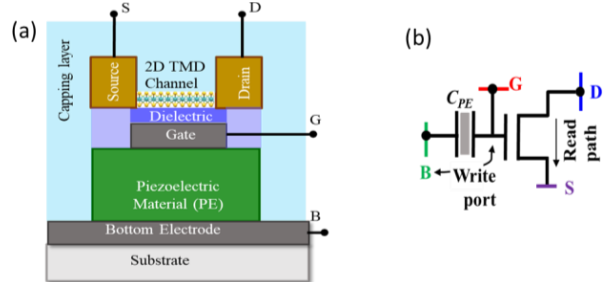

Fig. 1: (a) Device structure (b) schematic of PeFET.

Fig. 2: Write and read operation of PeFET.

## II. BACKGROUND

### A. Ferroelectric (FE) based NVMs

The ferroelectric RAM or FERAM is among the earliest NVM that utilized low power electric-field driven write. FERAM had garnered interest due to its highly compact 1 transistor-1 capacitor (1T1C) cell configuration and high retention (due to lower depolarization fields compared to FEFETs). However, a drawback of this design is destructive read which requires write-back and leads to read inefficiencies. On the other hand, FEFETs offering separate read-write paths overcome the issue of destructive read. Two flavors of FEFETs have been explored (a) with an inter-layer metal (ILM) between FE and the underlying transistor (also called FEMFET) and (b) without an ILM i.e. with FE integrated directly on the channel or the gate dielectric (DE). FEFET without ILM features polarization-induced threshold voltage shift, which can be utilized to sense the stored data. While offering useful features such as multi-level storage and compact 1T design [13], FEFETs without ILM need high write voltage and suffer from variability, endurance and retention concerns due to traps at the FE-DE interface and depolarization fields in the FE. On the other hand, FEFETs with ILM exhibit mitigation of some of these issues because (a) the ILM decouples the cross-sectional optimization of FE and the underlying transistor, which helps in write voltage reduction [8] and (b) the ILM between FE and DE reduces traps and related issues. However, the presence of a floating ILM makes this device vulnerable to gate leakage, which leads to severe bit sensing challenges [9]. Furthermore, NVMs based on FEFETs with ILM typically require one or more access transistors (e.g. 2T/3T/4T designs [22]) to selectively access cells in an array, reducing their integration density. In the proposed PeFETs, we attempt to overcome some of the cited limitations of FE-based NVMs while exploiting low-power *E*-driven write in FE. As described later, the unique piezoelectricity-induced bandgap modulation based read in PeFETs enables a four-terminal device design with gate, drain, source and back contacts. They serve as knobs to discretely control read-write paths, enabling non-destructive read. The absence of floating metal in this device eliminates gate leakage concerns. Further, the PE layer is controlled by metals on both its sides, alleviating the issues of traps, variability and retention.

### B. Piezoelectric (PE) based devices

An important design consideration for PeFET is the choice of PE material. We envision utilization of both ferroelectric and piezoelectric properties of PE material in this device. To this end, PZT-5H is one of the suitable materials as it possesses (i) sufficiently wide hysteresis of polarization-voltage response for non-volatility and (ii) large enough strain for achieving effective bandgap modulation in 2D-TMD. This is different from the previous proposals of PE based FETs, e.g., PieozoElectronic Transistor (PET) [14] and Electrostrictive FET (EFET) [10] that were designed as steep-switching/logic devices. Hence, large piezoelectric coefficient is the primary requirement for the PE material in these devices. They utilize PMN-PT to modulate the resistance of piezoresistive material in PET [14] or bandgap of 2D-TMD channel in EFET [10]. However, PMN-PT is not ideal for the proposed NVM due to its low coercive field despite its impressive piezoelectricity.

Another important design aspect of PeFET is that the 2D-TMD in the channel should allow sufficient piezoelectricity induced bandgap modulation. Various experiments have demonstrated monotonic bandgap reduction in 2D-TMD, especially $MoS_2$, on application of out-of-plane pressure. For example, monolayer $MoS_2$ achieves bandgap reduction of up to ~800meV/GPa [11]. EFET proposed in [10] discusses modulation of $MoS_2$ with piezoelectricity induced out-of-plane pressure in the context of logic applications. We extend the idea of piezoelectricity driven bandgap modulation of 2D-TMD to NVM design [15]. Our PeFET NVM stores bit information in PE and leverages polarization-dependent piezoelectric response to modulate the bandgap of mono-layer $MoS_2$ for sensing.

## III. PIEZOELECTRIC FET (PeFET) DESIGN AND MODELING

### A. Device structure and operation

PeFET consists of four terminals, namely drain (*D*), gate (*G*), source (*S*) and back (*B*) contacts (Fig. 1). A piezoelectric (PE) material (which also exhibits ferroelectricity) sits between *G* and *B*. The bit is stored in the form of polarization (*P*) in PE, with +(-) *P* representing logic '1' ('0') (Fig. 2). Application of voltage across *G* and *B* ($V_{GB}$) enables writing states 1/0 to PE by switching *P*. $V_{GB}$ greater than coercive voltage of PE ($V_C$) leads to +*P* in the PE, while $V_{GB}$ < -$V_C$ yields -*P* storage.

In addition to write, *G* plays an important role in sensing of the PeFET states. It is used to generate *P*-dependent piezoelectric behavior in PZT. Moreover, it electrostatically controls the 2D-TMD channel (monolayer $MoS_2$). A positive read voltage ($V_R$) is applied to *G* such that $0 < V_R < V_C$. Such $V_R$ creates strain in PE ($S_{PE}$) [16] by modulating its thickness ($\Delta t_{PE}$) based on the stored *P* (Fig. 2). For +*P*, PE thickness extends, i.e., $\Delta t_{PE} > 0$ and $S_{PE} > 0$; while -*P* results in shrinking of PE thickness, $\Delta t_{PE} < 0$ and $S_{PE} < 0$ (Fig. 2). Consequently, $S_{PE}$ yields a positive or negative stress in PE ($\sigma_{PE}$). $\sigma_{PE}$ transduces as pressure in TMD ($\sigma_{TMD}$). $\sigma_{TMD}$ causes dynamic modulation in the bandgap ($\Delta E_G$) of TMD [11] (Fig. 2). If $\sigma_{TMD}$ > 0 (resulting from $S_{PE} > 0$ or +*P*), $\Delta E_G < 0$ and bandgap reduces. Otherwise, when $\sigma_{TMD} < 0$ (for $S_{PE} < 0$ or -P), $\Delta E_G > 0$ and bandgap is higher. The effect of $\Delta E_G$ reflects in drain current ($I_{DS}$) as low/high resistance states (LRS/HRS) (Fig. 2). For +P, $S_{PE} > 0$, $\Delta E_G < 0$ and $I_{DS} = I_{LRS}$ (high) and -P yields $S_{PE} < 0$, $\Delta E_G > 0$ and low $I_{DS} = I_{HRS}$ (Fig. 5(b)).

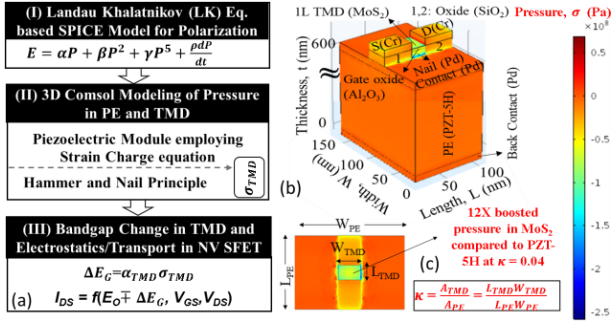

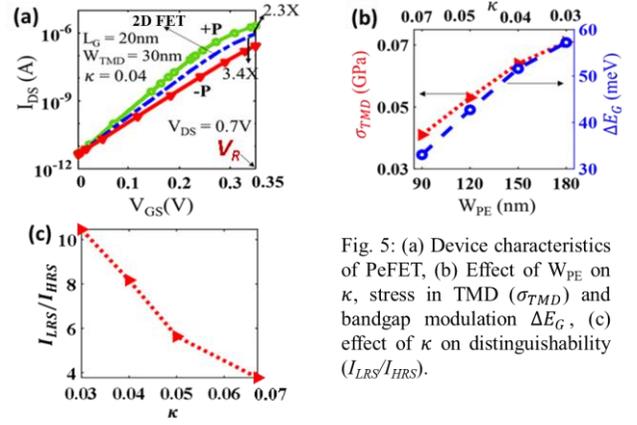

Fig. 3: (a) Simulation framework of PeFET, (b) 3D Modeling of PeFET in COMSOL Multiphysics, (c) Effect of 'hammer-and-nail' on PeFET showing boosted pressure in TMD.

Fig. 5: (a) Device characteristics of PeFET, (b) Effect of $W_{PE}$ on $\kappa$, stress in TMD ($\sigma_{TMD}$) and bandgap modulation $\Delta E_G$, (c) effect of $\kappa$ on distinguishability ($I_{LRS}/I_{HRS}$).

To efficiently transduce $\sigma_{PE}$ to $\sigma_{TMD}$, we utilize hammer and nail effect, similar to [14], wherein the area of 2D-TMD above the gate ($A_{TMD}$) acts as the nail while PE serves as the hammer (Fig. 3(b)). We design $A_{TMD}$ to be smaller than the area of PE ($A_{PE}$) by choosing appropriate PE dimensions. Note that $A_{TMD}$ is fixed by the feature size (F) and width (W) of PeFET. In our design, we choose F=20nm and W=30nm (minimum width for high integration density). We select metals with high stiffness for the nail, PE and source/drain contacts (e.g., Pd, Cr) and encapsulant surrounding the device to minimize loss of $\sigma_{PE}$. The purpose of the encapsulant is to provide favorable boundary conditions for stress to be maximized at the TMD channel.

### B. Modeling

We develop a simulation framework shown in Fig. 3 to analyze the proposed PeFET. We model *P-E* response of PZT-5H using Landau-Khalatnikov (LK) equation [17] and calibrate it with experimental data in [16]. Further, we model pressure transduced from PE to 2D-TMD ($\sigma_{TMD}$) using COMSOL Multiphysics Suite. We simulate 3D structure of PeFET (including hammer and nail effect) in COMSOL and employ strain-charge form of the constitutive equations for PE to obtain $\sigma_{PE}$ and $\sigma_{TMD}$. We use boundary conditions on the top surface of the device (over the contacts and TMD) that represent a stiff encapsulant material. The piezoelectric coefficients ($d_{33}$ and $d_{31}$) of PZT-5H for this simulation are extracted from experimentally obtained *S-E* response in [16]. $\sigma_{TMD}$ is converted to $\Delta E_G$ using reported bandgap coefficient ($\alpha$) for mono-layer MoS$_2$ [11]. $\Delta E_G$ is self-consistently coupled with Verilog-A based 2D-TMD FET model [18] by accounting for impact of dynamically changing bandgap on charge/potential of 2D-TMD channel. The 2D-TMD FET model provides *P*-dependent $I_{DS}$ of PeFET, and is used in HSPICE to design the NVMs. The parameters used are based on experiments or reported literature [14], [16], [19], [20] and summarized in Fig. 4.

| | | |
|---|---|---|
| $\alpha$ = -3.95e6m/F | $\beta$ =1.26e6m$^5$/F/C$^2$ | $\gamma$= 3.21e8m$^9$/F/C$^4$ |
| $d_{33}$= 650pm/V | $d_{31}$= -320pm/V | $\mu_{TMD}$ = 90cm$^2$/Vs |
| $R_C$= 200Ω-μm | $\alpha_{TMD}$= 806eVGPa$^{-1}$ | $E_0$ = 1.5eV |
| $t_{PE}$ = 600nm | $t_{nail}$= 10nm | $t_{ox}$= 3nm |
| $t_{TMD}$ = 0.65nm | $L_G$ = 20nm | $W_{TMD}$= 30nm |

Fig 4: Parameters used for simulation

### C. Device characteristics

Let us start by understanding the strain transduction characteristics of the PeFET. Our results from COMSOL simulations (Fig. 3(b)) show that the hammer and nail effect boost $\sigma_{TMD}$ compared to $\sigma_{PE}$, when the area of nail/2D-TMD ($A_{TMD}$) is smaller than that of PE ($A_{PE}$), the hammer. The device parameter $\kappa = A_{TMD}/A_{PE}$ is a measure of this effect [14], where smaller $\kappa$ is expected to provide larger $\sigma_{TMD}$. We observe ~12X increase in $\sigma_{TMD}$ compared to $\sigma_{PE}$ at $\kappa$ = 0.04 (Fig. 3(c)). We will explain the effect of $\kappa$ in more detail in the next sub-section.

Let us now discuss polarization/strain-dependent transfer characteristics ($I_{DS}$-$V_{GS}$) of PeFET (Fig. 5(a)) with $\kappa = 0.04$. To avoid *P*-switching, we apply gate voltage ($V_G$) < $V_C$ of PZT-5H (= 0.6V at PE thickness of 600nm), while keeping back voltage ($V_B$) = 0. For comparison, we also simulate a baseline device with $V_{GB}$=0 (i.e. with $V_G$ and $V_B$ swept together), from which we obtain *P*-independent nominal transfer characteristics of MoS$_2$-FET. Our results show at $V_{GS}$ = 0.35V, +*P* state in the PeFET yields $\sigma_{TMD}$ = 0.64GPa and $\Delta E_G$ = 51 meV causing 2.3X higher $I_{DS}$ ($I_{LRS}$). While, -*P* results in 3.4X lower $I_{DS}$ ($I_{HRS}$) at $V_{GS}$ = 0.35V compared to baseline.

Based on $I_{DS}$-$V_{GS}$ characteristics, we identify that 0.3V < $V_{GS}$ < 0.4V provides optimal distinguishability ($I_{LRS}/I_{HRS}$), necessary current for read operation and ample read disturb margin ($V_C$ - $V_{GS}$ ~ 200mV). We choose $V_{GS}$ = $V_R$ = 0.35V which leads to $I_{LRS}/I_{HRS}$ ~ 8X (Fig. 5(c)) at $\kappa$ = 0.04. Note that $I_{LRS}/I_{HRS}$ can be improved by device optimization (e.g. by reducing κ), which we discuss in the next sub-section.

### D. κ analysis

Lower $\kappa$ boosts stress in TMD due to '*hammer and nail effect*'. $\kappa$ tuning is achieved by appropriately choosing the PE width ($W_{PE}$), and thereby $A_{PE}$ (Fig. 5(b)). This leads to a design time optimization of $\sigma_{TMD}$ and $\Delta E_G$. By increasing $W_{PE}$ from 90nm to 180nm, $\kappa$ decreases from 0.07 to 0.03, leading to 1.78X increase in $\sigma_{TMD}$ and $\Delta E_G$ (Fig. 5(b)). Hence, for +*P*, more aggressive bandgap reduction is observed at small $\kappa$, increasing $I_{LRS}$. Whereas in -*P*, bandgap increases with a decrease in $\kappa$, which weakens $I_{HRS}$. Strong $I_{LRS}$ combined with weak $I_{HRS}$ increases $I_{LRS}/I_{HRS}$ and hence distinguishability from 3X to 11X as $\kappa$ is reduced from 0.07 – 0.03 (Fig. 5(c)).

## IV. PeFET Based Non-Volatile Memory Designs

With the understanding of the device operation of PeFETs, we now propose three flavors of PeFET-based NVMs, namely, high density (HD), 1T-1PeFET and cross-coupled (CC) cells.

### A. High density (HD) Non-Volatile Memory

HD NVM is designed with 1-PeFET, without an access-transistor (Fig. 6(a, b)). *G* of PeFET is connected to word line (*WL*) that is shared along a row. *B* and *D* are connected to write bit-line (*WBL*) and read bit-line (*RBL*) respectively, (Fig. 6(c)) and are shared along a column. Notice that *B* is being shared by adjacent cells in a column (Fig 6(b)) which helps in minimizing area. Next, we discuss the write and read operations.

*1. Write:* *WL* and *WBL*, connected in cross-point fashion, form the write port with $C_{PE}$ between them [Fig. 6(c)]. To write, *WBL* and *WL* of the accessed cells are asserted such that $|V_{GB}| = V_{DD} > V_C$ appears across PE, resulting in *P* switching. We implement a two phase write operation to enable writing the states '1' and '0' in different cells of a word (in the same row). To write '1', we drive *WBL* to -$V_{DD}$/2. Next, we drive *WL* to -$V_{DD}$/2 in phase one ($\Phi_1$) and switch to $V_{DD}$/2 in the second ($\Phi_2$). Consequently, $V_{GB} = V_{WL} - V_{WBL} = 0$ in $\Phi_1$ which has no effect on *P* and $V_{GB} = V_{WL} - V_{WBL} = V_{DD}$ causes PE to switch to +*P* ('1') in $\Phi_2$. Similarly, '0' is written by driving *WBL* to $V_{DD}$/2 and asserting *WL* like before. Now, $V_{GB} = V_{WL} - V_{WBL} = -V_{DD}$ switches polarization to -*P* in $\Phi_1$ and $V_{GB} = 0$ retains -*P* in $\Phi_2$.

With write using $\pm V_{DD}$/2 for accessed cells, *WBLs*/*WLs* of half-accessed/un-accessed cells can be kept at 0V. Hence, half-accessed cells experience $|V_{GB}| \leq V_{DD}/2 < V_C$, and un-accessed cells have $V_{GB} = 0$, thereby averting write disturbs. Having *BL*/*WL* of half-accessed/un-accessed at 0V cuts down write energy that may be present in other cross-point architectures based on $V_{DD}$/2 or $V_{DD}$/3 biasing for half/un-accessed cells [21].

Note that due to absence of access transistors (*AX*), use of negative voltages for write does not incur large energy overheads compared to previous designs based on other technologies involving *AX* [22]. In latter designs, negative write voltage also needs to ensure *WLs* of *all* the unaccessed rows are driven to negative voltage to avert their turning ON. This incurs a significant energy cost, which is absent in HD design. For the same reason, in our other PeFET designs that employ AX, we constrain write biasing voltage to be positive (details later).

*2. Read:* We apply $V_R (= V_{DD}/2 < V_C)$ on *WL* and $V_{DD}$ to *RBL* of accessed cells. Read current is sensed on *RBL* in a single phase. Half accessed rows have *WL* at 0V while in half accessed columns, *RBL* voltage is 0V. Hence, current flow through half-accessed/unassessed cells is negligible or zero.

The lack of *AX* in HD cells and sharing of back contact in the layout results in a compact design. However, absence of *AX*

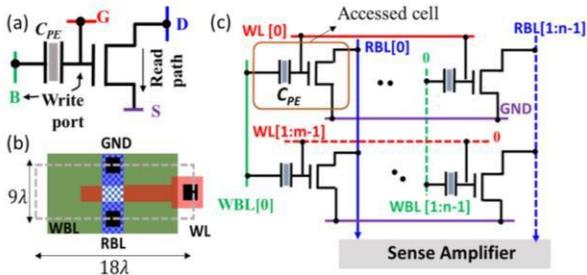

Fig. 6 (a) Schematic (b) Layout and (c) Array of HD PeFET

exposes PE capacitance ($C_{PE}$) of half-accessed cells directly on *WBL*. Since $C_{PE}$ is typically large (dielectric constant ~ $4000\epsilon_0$), write energy efficiency is adversely affected. Based on our analysis, we attribute that 78% of write energy is originated from charging/discharging of $C_{PE}$, followed by 12% due to metal capacitance switching of *BLs*/*WL* and 10% due to *P*-switching. To minimize effect of $C_{PE}$ on write energy, we propose PeFET NVMs with *AX* that isolates *WBL* from $C_{PE}$.

### B. 1T-1PeFET Non-Volatile Memory [Fig.7(a)]

1T-1PeFET consists of an n-type 2D FET connected to back contact (*B*) of a PeFET. Notice that unlike in HD NVM, large $C_{PE}$ on *WBL* is replaced by smaller drain capacitance from *AX* in a column. We explore two variants of layout for this cell: (i) tall layout for area optimized design [Fig. 7(b)] and (ii) wide layout where the cell height is minimized [Fig. 7(c)]. The latter is motivated from our understanding that bit line capacitance is the dominating contributor for delay and energy in 1T-1PeFET arrays. For example, 79% of its total write energy is attributed to $V_{DD}/2$ switching of *WBLs* in half-accessed cells. *WBLs* of half-accessed cells must be driven to $V_{DD}/2$ to ensure $|V_{GB}| < V_C$ across their PEs in order to prevent *P*-switching. Such a requirement also causes a large overhead in write energy. To mitigate this, we present Segmented 1T-1PeFET NVM.

### C. Segmented Architecture of 1T-1PeFET NVM

1T-1PeFET segmented architecture [Fig. 7(e)] which eliminates energy overheads from half-accessed/unassessed cells is motivated by FERAM architecture [23].

The 1T-1PeFET segmented array is divided into multiple segments each comprising of 64 columns that correspond to size of a word ($N_W$) and $N_R$ rows. If the entire array is constituted of $N_C$ columns and $N_R$ rows, there are $N (= N_C/N_W)$ segments, each of size $N_W$ x $N_R$. Moreover, global plate line (*GPL*) is responsible for the said segmentation. The array contains $N = N_C/N_W$ *GPLs*. *GPLs* provides input to $N_R$ buffers in each segment. These buffers act as gates for a segment and their output drives a local plate line (*LPL*). *LPL* connects the 64 cells (= $N_W$). Note that *LPL* is loaded with a smaller net $C_{PE}$ from 64 cells instead of an entire row (as in a standard architecture). The *WL*, which runs through a row, provides supply voltage to the buffers and also activates the *AX*s of 1T-1PeFET NVMs in the accessed row. In addition, each column in a segment has its own *WBL* and *RBL*. In summary, each segment comprises of (i) a *GPL*, (ii) $N_R$ *LPLs*, (iii) $N_R$ *WLs* that are shared by other segments, (iv) $N_W$ *WBLs* and (v) $N_W$ *RBLs*. Now, let us understand the implication of the segmented architecture on the write and read operations of 1T-1PeFET.

*1. Write:* Two phase signal, i.e., 0 ($\Phi_1$) → $V_{DD}$ ($\Phi_2$) is applied to *GPL* of the accessed segment during write. Write input data $0/V_{DD}$ is provided to $N_W$ *WBLs*. Let the first word of first segment (accessed word in Segment [0] of Fig. 7(e)) is to be written to, without any loss in generality. *WL* is asserted with $V_{DD} + V_{TH}$ which activates Buffer[0]. *LPL* of the accessed word ($LPL_0[0]$) is 0 and $V_{DD}+V_{TH}$ when *GPL* is 0 and $V_{DD}$ respectively. If *WBL* = 0, $V_{GB} = 0$ in $\Phi_1$ and $V_{GB} = V_{DD} + V_{TH}$ in $\Phi_2$. Hence, +*P* state (logic 1) is written in $\Phi_2$. Similarly, if *WBL* = $V_{DD}$, $V_{GB}$ = -$V_{DD}$ in $\Phi_1$ and -*P* state (logic 0) is being written. All *RBLs* along with *WBLs* and *GPLs* of un-accessed words/segments (e.g., *GPL*[1] to *GPL*[N-1]) are kept at 0V.

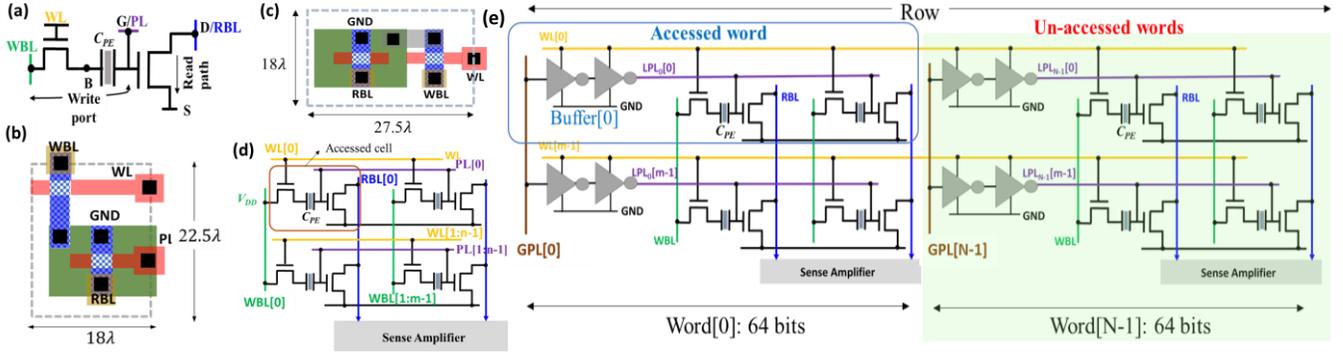

Fig. 7 (a) Schematic (b) tall layout (c) wide layout, (d) array and (e) segmented array of 1T-1PeFET

Hence, their *LPLs* are also at 0V. Thus, energy expended by these segments is minimized in this architecture.

*2. Read:* We accomplish read by driving *GPL* of accessed segment and *WL* to $V_R$ while keeping *WBLs* at 0. This results in $V_R$ on *LPL* (e.g., $LPL_0[0]$ in Fig. 7(e)). Read current is sensed with *RBL* voltage at $V_{DD}$. The voltages of *GPL*, *WBL*, *RBL* of un-accessed sections and *WL* of un-accessed rows are 0V.

### D. Cross-coupled PeFET NVM (CC) [Fig. 8(a)]

In 1T-1PeFET, we focused on design aspects that minimized impact of $C_{PE}$ on cell performance and energy consumption. By applying segmentation, we further minimized the interaction of $C_{PE}$. However, a limitation of 1T-1PeFET is the area overhead added by *AX*. Also, unlike in HD cell, the back contact is an internal node in 1T-1PeFET that cannot be shared, constraining its compactness as per design rules. Furthermore, although reduced considerably, $C_{PE}$ still needs to be switching during read. To alleviate the burden of $C_{PE}$ during read while also targeting cell area reduction compared to 1T-1PeFET cells, we propose cross-coupled (CC) PeFET NVM cell. The technique of cross-coupling two PeFETs in this cell allows us to share the back-contact (along a row), thereby conserving cell area. Our layout analysis [Fig. 8(b)] suggests that the area per bit in CC is 1.36X and 1.66X lower than 1T-1PeFET tall and wide cell respectively at $\kappa = 0.04$. Moreover, the back contact (hence $C_{PE}$) need not be switched in CC NVM during read (remains at 0V), thus improving read latency and energy.

We design CC cell [Fig. 8(a)] with two PeFETs (A and B) that are cross-coupled by connecting *D* of A ($D_A$) to *G* of B ($G_B$) and vice-versa. $D_A$ and $D_B$ are connected to bit-lines $BL_1$ and $BL_2$ respectively through $AX_1$ and $AX_2$. *G* of $AX_1/AX_2$ are controlled by *WL*. Back contacts of A and B are driven by *PL*. Bits in A and B can be stored, read and written individually. Hence, each CC cell is equivalent to a 2-bit NVM.

### E. Segmented Architecture of CC NVM (Fig. 9)

We propose segmentation in CC NVM array similar to 1T-1PeFET array so that *PL* has reduced $C_{PE}$ load during write. Each segment comprises of (i) $N_R$ rows, (ii) 32 columns of CC cells that effectively act as 64 bits of a word ($N_W$), (iii) a *GPL* that provides input to $N_R$ buffers, that are in turn activated by a *WL* and (iv) $N_R$ *LPLs* driven by the output of buffers. Notice that a *LPL* is a *PL* (back contact) shared horizontally by 32 CC cells within a segment. One key difference between 1T-1PeFET and CC segmented array is that the *LPL* voltage during read needs to be switched to $V_R$ in the former, while it is 0V in CC NVM. This eliminates the effect of $C_{PE}$ during read for improved read latency and energy efficiency. Next, we explain the write and read operations in a segmented CC array.

*1. Write:* Here, we explain various cases of writing to A and B. When we intend to store *complementary states* to A and B, i.e., *+P* (or -*P*) to A and -*P* (or +*P*) to B, $BL_1$ and $BL_2$ are driven to 0V (or $V_{DD}$) and $V_{DD}$ (or 0V) respectively. 0 ($\Phi_1$)→$V_{DD}$ ($\Phi_2$) is applied to *GPL* of selected segment and *WL* is asserted with $V_{DD}+V_{TH}$. *LPL* receives 0→$V_{DD}$ as output. Moreover, voltage at $BL_1$ and $BL_2$ is passed to $G_B$ and $G_A$ respectively by CC NVM's *AXs* after *WL* assertion. For PeFET A, $V_{GB}$ of A ($V_{GB,A}$)=$V_{BL2}-V_{LPL}$. Similarly, $V_{GB,B}$ ($V_{GB}$ of B) is $V_{BL1}-V_{LPL}$. Say, $V_{BL1}$=0, $V_{BL2}$=$V_{DD}$ and *LPL* is signaled with 0 ($\Phi_1$) → $V_{DD}$ ($\Phi_2$), then we get $V_{GB,A}$=$V_{BL2}-V_{LPL}$=$V_{DD}$ in $\Phi_1$; +*P* is being written to A. For B, $V_{GB,B}$=$V_{BL1}-V_{PL}$=-$V_{DD}$ during $\Phi_2$ which is when -*P* is written.

To store -*P* in both A and B, $BL_1$ and $BL_2$ are maintained at 0V. After *WL* is asserted and 0→$V_{DD}$ appears on *LPL* (like before), $V_{GB,A}/V_{GB,B}$=-$V_{DD}$ is obtained during $\Phi_2$ switching both PEs to -*P*. To write +*P* to both A and B, $BL_1$ and $BL_2$ are driven to $V_{DD}$. In this context, an important aspect needs to be highlighted. When *WL* is asserted, voltage from $BL_1$ and $BL_2$ (which are at $V_{DD}$) charge $D_A$ (and $G_B$) and $D_B$ (and $G_A$). As a result, PeFETs A and B are turned on, while write ensues. This causes $D_A$ and $D_B$ (and connected $G_B$ and $G_A$) to drop to $V_{DD}$ - $\Delta V$. Reduced gate voltage implies lower $V_{GB}$ across PEs of both PeFETs, which ultimately increase write latency.

*2. Read:* Currents through $BL_1$ and $BL_2$ are sensed after driving them to $V_R$. *WL* is asserted with $V_{DD}$ so that both cells get $V_G = V_R$ required for read. *GPL* and hence *LPL* of accessed segment is at 0V during read. The idea is to have $V_{GB} = \sim V_R$ for both A and B. Depending on whether +*P* or -*P* is stored in A (or B), $I_{LRS}$ or $I_{HRS}$ is sensed on $BL_1$ (or $BL_2$). As in write, cross-coupling has important effects during read which we elaborate next.

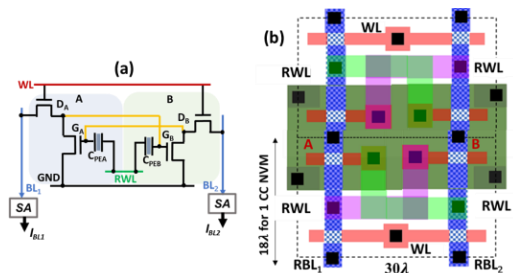

Fig. 8: (a) Schematic and (b) layout of CC NVM.

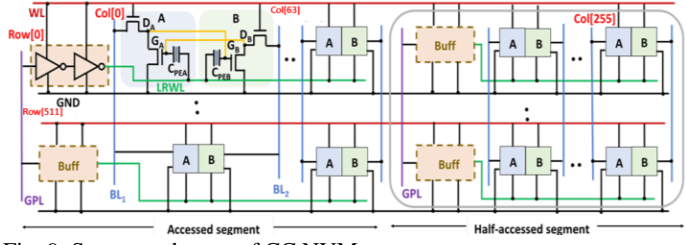

Fig. 9: Segmented array of CC NVM

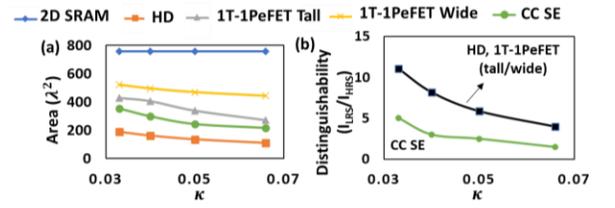

Fig. 11 (a) Area and (b) distinguishability: varying $\kappa$

a. Let, A/B store $-P/-P$. Since both cells are in HRS, $D_A$ and $D_B$ in Fig. 9 is at $V_R$. Let currents sensed on $BL_1/BL_2$ be $I_{HRS00}$.

b. Next, let $-P/+P$ be the stored states of A/B respectively. As A is in *HRS*, $D_A$ is pulled to $V_R$ as desired. On the other hand, the $D_B$ is pulled down to $V_R - \Delta V$ by action of *LRS* in cell B. $G_A$ and $G_B$ have similar voltage as $D_B$ and $D_A$ due to cross-coupled connection. So, cell A gets a $V_{GB, A} = V_R - \Delta V < V_R$. This lowered gate voltage makes A (storing $-P$) more resistive than HRS in $-P/-P$ state (previous example) and now we obtain $I_{HRS01}$ on $BL_1$ which is lesser than $I_{HRS00}$ in (a). For B, $G_B$ is $V_R$ and we obtain $I_{LRS01}$ (with $V_{GB}=V_R$ and $V_{DS}=V_R-\Delta V$) on $BL_2$. If A/B has $+P/-P$ instead, we obtain $I_{HRS10}$ on $BL_1$ and $I_{LRS10}$ on $BL_2$.

c. When $+P$ is stored in both PeFETs (i.e., they are in LRS), their drain and connected gate charge to $V_R - \Delta V$. Hence, $V_{GB, A/B} = V_{DS, A/B} = V_R - \Delta V$. Diminished $V_{GB}$ allows lower current (let it be called $I_{LRS11}$) than $I_{LRS10}$ or $I_{LRS01}$ on $BL_1$ and $BL_2$ in this LRS state. Finally, we conclude that $I_{LRS11} < I_{LRS10}$ and $I_{HRS00} > I_{HRS10}$ or $I_{HRS01}$. We define distinguishability for CC NVM, considering the worst case, as the ratio - $I_{LRS11}/I_{HRS00}$, which is lower than the other PeFET NVMs. Note that in HD and 1T-1PeFET, distinguishability is identical to $I_{LRS10}/I_{HRS10}$.

While CC NVM exhibits overheads and complex read/write due to cross-coupling, its advantage (as noted earlier) is effective contact sharing lead lower bit area than 1T-1PeFET. Next, we quantify area, performance, energy efficiency of the proposed PeFET NVMs and compare them to 2D-FET SRAMs.

## V. ANALYSIS OF PeFET BASED NVM ARRAYS ($\kappa = 0.04$)

### A. Layout and area [Fig. 10 (a)]

We use $\lambda$ (= Gate length/2) based rules and gate/metal pitch defined by Intel 20nm process [24] for layout analysis of PeFET NVMs. In HD NVM, back terminal is shared among adjacent cells in a column (*WBL*) resulting in a cell height ($H$) of one poly-pitch (PP = $9\lambda$). Cell width ($W = 18\lambda$) is controlled by width of PE (dictated by $\kappa$ as discussed in Section III). Finally, an HD cell's area is $162\lambda^2$. We present tall [Fig. 7(a)] and wide [Fig. 7(b)] layouts for 1T-1PeFET NVM. In the tall variant, we minimize cell area by stacking *AX* along *H*. *WBL* contact (or drain of *AX*) is shared with adjacent cells in the column. However, sharing of *RBL* contact is restricted by the internally positioned back terminal. Hence, $H$ of 1T-1PeFET tall cell is equal to 2.5 PP (22.5$\lambda$) or 2.5X larger footprint than HD. Moreover, we propose wide layout (1T-1PeFET wide NVM) with decreased $H = 2$ PP (18$\lambda$) to minimize *BL* capacitance for improved energy/latency. The $\kappa$-dependent width of the back contact and access transistor width including contacts sets *W*. Its area shows an overhead of 3.1X over HD. Area of CC NVM is 1.4X and 1.7X lower than 1T-1PeFET tall and wide NVM respectively, while it is 1.8X larger than HD cell. Improvement of CC NVM area over 1T-1PeFET is achieved by efficient contact sharing (a) of back contact across *PL* (along rows unlike in other cells) and (b) of *BL* with adjacent cells (compare that with limited contact sharing in 1T-1PeFETs). HD, CC, 1T-1PeFET tall and wide NVMs are 4.7X, 2.5X, 1.87X and 1.53X area efficient compared to 2D-FET SRAM.

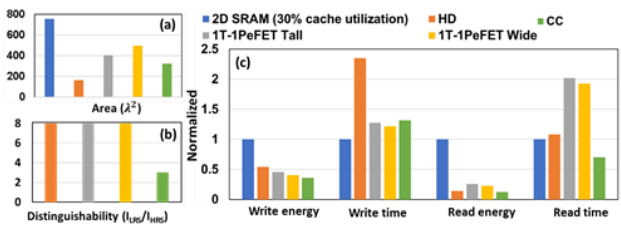

Fig. 10: Analysis of PeEFT NVMs compared to 2D-FET SRAM.

### B. Distinguishability [Fig. 10 (b)]

HD and 1T-1PeFET exhibit same distinguishability ($I_{LRS}/I_{HRS}$ = 8X) due to similar strain transduced at a fixed $\kappa$. In CC NVM, lower distinguishability of 3X is observed at same $\kappa$ due to limiting read condition - $I_{LRS10}/I_{HRS10}$ (refer Section IV(E2c).

### C. Write energy and latency [Fig. 10 (c)]

HD NVM shows degraded write energy and latency than other designs due to $C_{PE}$ from half-accessed cells adding to *WBL* and *PL* (refer Section IV). In Segmented 1T-1PeFET NVM, *AX* shields *WBL* from this capacitance leading to energy and latency improvements over HD. 1T-1PeFET wide design shows minimum latency due to its layout-enabled minimization of *BL* capacitance. CC NVM combines the benefits of *AX* (like in 1T-1PeFET) with a smaller area for further write energy optimization. However, it shows higher latency than the 1T-1PeFETs despite its smaller cell area due to *P*-switching occurring at diminished $V_{GB}$ (described in Section IV). Finally, write energy of HD, Segmented 1T-1PeFET tall, wide and CC NVM is 48%, 56%, 61%, 65% lower than 2D-FET SRAM. Note that these benefits are reported considering 30% utilization [26] for L2 cache. The baseline SRAM cache leaks for the remaining 70% utilization adding leakage energy, whereas PeFET NVMs do not. Large size of SRAM cell along with leakage energy are contributors for its high write energy. Moreover, we observe 22%, 28%, 31% and 134% overhead in latency in 1T-1PeFET wide, tall, CC and HD NVMs (caused by polarization switching) compared to 2D-FET SRAM.

### D. Read energy and latency [Fig. 10 (c)]

Among PeFET designs, CC shows minimum read energy as it is compact and *PL* connected to $C_{PE}$ is not activated during read. HD NVM shows improved read energy than 1T-1PeFET NVMs due smaller *RBL* capacitance (due to lower layout height) compared to other two designs. Since *RBL* switching energy is a dominant component of read energy, its reduction affects read energy similarly. We observe lower read energy in PeFET designs than 2D-FET SRAM attributed to (i) negligible leakage and (ii) capability to selectively turn of *RBL* discharge of half-accessed cells during read in PeFET arrays. Note that we access 64 bits during read/write operations for both PeFET and SRAM based arrays for consistency of comparison. Read energy of CC, HD, 1T-1PeFET wide and tall shows 87%, 85%, 77% and 74% improvement respectively compared to 2D-FET SRAM at 30% cache utilization. Impact of $C_{PE}$ during read causes latency to increase in HD, 1T-1PeFET tall and 1T-1PeFET wide by 7%, 93% and 100% compared to 2D-FET SRAM. Contrarily, CC cell shows 30% improvement since $C_{PE}$ is bypassed during read (by keeping *LPL* to 0V).

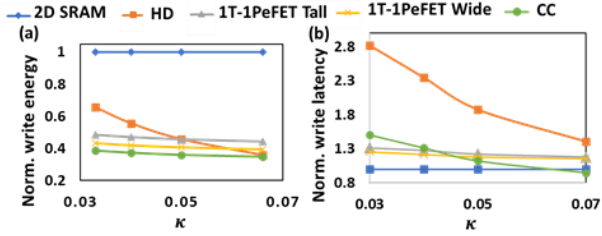

Fig. 12 Write (a) energy and (b) latency: varying $\kappa$.

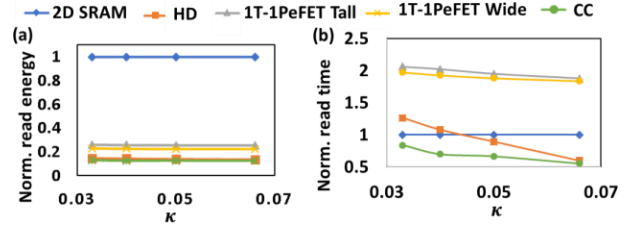

Fig. 13 Read (a) energy and (b) latency: varying $\kappa$.

## VI. $\kappa$ ANALYSIS OF PeFET BASED NVM ARRAYS

We have established in Section III that small $\kappa$ improves strain transduction to 2D-TMD channel, however at the cost of device footprint. Here, we discuss the effect of $\kappa$ in NVM array design.

### A. Area [Fig. 11(a)]

Increasing $\kappa$ leads to decrease in cell area of PeFET NVMs. With respect to 2D-FET SRAM, area improvement for PeFET NVM HD ranges from 4X at $\kappa = 0.03$ and to 7X at $\kappa = 0.07$. Size of CC cell is 2.15X-3.5X lower than 2D-FET SRAM for $\kappa = 0.03 - 0.07$. Whereas, 1PeFET tall and wide cells are 1.76X-2.8X and 1.45X-1.71X smaller for $\kappa = 0.03 - 0.07$.

### B. Distinguishability [Fig. 11(b)]

The efficiency of strain transduction increases at a lower $\kappa$, which boosts distinguishability. For $\kappa$=0.03, it is 11X compared to 3X at $\kappa$=0.07 in HD, 1T-1PeFET tall and wide designs. In CC, it lies in the range of 1.5X – 5X for $\kappa = 0.07 – 0.03$.

### C. Write energy and latency [Fig. 12(a, b)]

Write energy and latency of PeFET NVMs improves at higher $\kappa$ (lower PeFET footprint). Large $\kappa$ decreases (i) *WL* metal capacitance, and (ii) FE/PE switching capacitance in all PeFET NVMs. In HD, it also reduces $C_{PE}$ of half-accessed cells that adds to *WBL*/*WL*. Hence, HD exhibits a steeper change in energy/latency with $\kappa$. Finally, HD, 1T-1PeFET tall, wide and CC NVM shows up to 36% - 65%, 52% - 56%, 58% - 61% and 62% - 66% improvement in write energy compared to 2D-FET SRAM respectively for $\kappa = 0.03 - 0.07$ (Fig. 12(a)). Write latency overhead over SRAM decreases from 181% (at $\kappa = 0.03$) to 40% (at $\kappa = 0.07$) for HD, 31%-17% for 1T-1PeFET tall, 26%-15% 1T-1PeFET wide designs (Fig. 12 (b)). CC cell shows an improvement of 5% over 2D-FET SRAM at $\kappa = 0.07$.

### D. Read energy and latency [Fig. 13(a, b)]

Read energy of all PeFET designs are nearly insensitive to $\kappa$. This is because read energy is dominated by *BL* capacitance switching energy, which is unaffected by $\kappa$ [Fig. 13 (a)]. Note by changing $\kappa$, we do not change the layout height and hence *RBL* capacitance; we only change the horizontal dimension of the layout. Read latency decreases as $\kappa$ increases due to smaller cell size and reduced $C_{PE}$ associated with *WL/PL* in HD and 1T-1PeFET designs. HD and CC designs show up to 40% and 44% improvement for $\kappa = 0.07$, whereas the overhead in 1T-1PeFET tall and wide designs decrease to 87% and 83% respectively.

## VII. FUTURE OUTLOOK

We have previously elicited the trade-offs between distinguishability, energy and area in PeFET based NVMs. We further proposed device-circuit design techniques to mitigate them, such as, geometry related $\kappa$ optimization, use of access transistor to design 1T-1PeFET versus access-transistor-less HD, layout optimization of 1T-1PeFET and segmentation in 1T-1PeFET/CC arrays. In the future, we envision that optimization of PeFET based NVMs could be furthered by alternative piezoelectric/ferroelectric materials. For example, Si-doped $HfO_2$ is a promising replacement for PE material used here - PZT-5H. Recent experiments have pointed to reasonable piezoelectric effects in Si-doped $HfO_2$ [27]. Moreover, Si-doped $HfO_2$ ensures good CMOS-compatibility, improved PE thickness scalability, potentially enhanced performance and energy-efficiency of PeFET NVMs. However, information of certain necessary parameters such as elastic tensor components of Si-$HfO_2$ is currently limited. Advanced characterization for such parameters and understanding endurance limits is required to establish it as the PE material for future PeFETs.

## VIII. CONCLUSION

We propose piezoelectric strain FETs (PeFET) utilizing polarization-based bit storage in ferroelectric material, electric field driven write and piezoelectricity induced dynamic bandgap modulation of 2D-TMD channel for bit sensing. We present HD, 1T-1PeFET tall, wide and CC NVMs based on PeFETs. We analyze them at the array level applying device and circuit optimizations for improved area-energy-latency tradeoff. HD surpasses other PeFET NVMs in compactness with up to 7X smaller area than 2D-FET SRAM. CC NVM stands out with benefits in write energy (66%), read energy (87%) and read latency (44%). However, slower write and weakened distinguishability are its limitations. 1T-1PeFETs are fastest to write, albeit with an overhead over 2D-FET SRAM and show good distinguishability (up to 11X), just as HD.